\documentclass[global,twocolumn]{svjour}
\usepackage{graphicx}

\journalname{to appear in Applied Physics B, Special Issue ``Optical
Properties of Nanoparticles''}

\renewcommand{\d}{\mathrm{d}}

\begin{document}

\title{Collectivity in the optical response of small metal clusters}

\author{S.\ K\"ummel\inst{1} \and K.\ Andrae\inst{2} \and P.-G.\
Reinhard \inst{2} }

\institute{Department of Physics and Quantum Theory Group, Tulane
  University, New Orleans, Louisiana 70118, USA \and
Institut f\"ur Theoretische Physik, Universit\"at Erlangen, D-91077
  Erlangen, Germany}
\mail{skuemmel@tulane.edu}

\date{Received:  / Revised version: }

\maketitle

\begin{abstract}
The question whether the linear absorption spectra of metal clusters
can be interpreted as density oscillations (collective ``plasmons'')
or can only be understood as transitions between distinct molecular
states is still a matter of debate for clusters with only a few
electrons. We calculate the photoabsorption spectra of $\mathrm{Na}_2$
and $\mathrm{Na}_5^+$ comparing two different methods: quantum
fluid-dynamics and time-dependent density functional theory. The
changes in the electronic structure associated with particular
excitations are visualized in ``snapshots'' via transition
densities. Our analysis shows that even for the smallest clusters, the
observed excitations can be interpreted as intuitively understandable
density oscillations. For $\mathrm{Na}_5^+$, the importance of
self-interaction corrections to the adiabatic local density
approximation is demonstrated.


\end{abstract}

\section{Introduction}
\label{intro}

Among the earliest experiments providing insight into the electronic
structure of metal clusters were measurements of the linear
photoabsorption spectra \cite{neupabs,selby}. Today, still, they are
among the most powerful probes of a cluster's structure, and
understanding the effects of collectivity in absorption spectra in
general is also one prerequisite for understanding the nonlinear
regime that is being probed experimentally with increasing
sophistication \cite{meiwesbroer}. In particular for positively
charged sodium clusters, linear spectra have been measured for a broad range
of cluster sizes and temperatures \cite{charpabs}. Typically, the
experiments show a few strong absorption lines that exhaust most of
the oscillator strength. However, despite the fact that the
absorption spectra have been known for a long time, their theoretical
interpretation is still being discussed. Since sodium is the
nearly-free electron metal par excellence, allowing to study metallic
behavior in one of its purest forms, a lot of theoretical work over
the years has been devoted to photoabsorption in Na clusters
\cite{ep,brack89,lrpa,guet,schmidt,rubio,ci,sic,cheli,locpp,baerends}. From
these studies, two different and somewhat opposing points of view on
the interpretation of the observed resonances emerged. On the one
hand, small clusters can accurately be described in the language of
quantum chemistry, understanding the excitations as transitions
between distinct molecular (electronic) states. On the other hand, the
experiments found an early, intuitive interpretation in terms of
collective oscillations of the valence electron density against the
inert ionic background, similar to the plasmon in bulk metals or the
giant resonances in atomic nuclei. Since the strong delocalization of
the valence electrons that characterizes nearly free electron metals
is found even in the smallest Na clusters, it has been argued that the
second interpretation should also be applicable to small clusters.

A theoretically well founded \cite{tddft} and practically tested
\cite{ep,schmidt,rubio,sic,cheli,baerends} method for the theoretical
investigation of excitations in metal clusters is time dependent
density functional theory (TDDFT) at the level of the adiabatic,
time-dependent local density approximation (TDLDA). A refinement of
TDLDA corrects for the self-interaction error leading to the scheme of
a time-dependent self-interaction correction simplified by a global
averaging procedure (TDSIC) \cite{sic}. A somewhat simpler, yet
powerful, alternative is quantum fluid-dynamics. In an extension of
earlier works \cite{brack89,lrpa}, quantum fluid-dynamics in a local
current approximation (LCA) was recently derived making direct use
\cite{lca} of the ground-state energy functional of density functional
theory. In this work we are comparing these methods using two very
small clusters as test cases with a threefold aim: First, comparing
LCA to TDDFT for exactly the same system allows to judge on the
reliability of LCA results. Second, by comparing TDLDA to its
on-average self-interaction corrected counterpart, we check the impact
of self-interaction corrections on low-energy photoabsorption
spectra. Third, the combination of methods allows us to demonstrate
that, indeed, the experimental spectra for even the smallest clusters
can be interpreted as valence electron density oscillations, leading
to an intuitive understanding of the experimentally observed effects.

In section \ref{methods} our theoretical methods are reviewed. Section
\ref{results} presents the results for $\mathrm{Na}_2$ and
$\mathrm{Na}_5^+$, which are discussed and summarized in section
\ref{conclusions}.

\section{Theory}
\label{methods}

Starting point of our investigations is the usual ground-state energy
functional  
\begin{eqnarray}
\label{clfunctional}
  E[n; \{\mathbf{ R}\}]&=&T_\mathrm{s}[\{\varphi\}]+E_\mathrm{ xc}[n]
+\int n(\mathbf{ r})V_\mathrm{ ion}(\mathbf{ r; \{R\}}) \,\d^3 r 
\nonumber\\&&
   +
  \frac{e^2}{2}\int \int
  \frac{n{\mathbf{(r)}n(\mathbf{ r'})}}{\left|\mathbf{ r-r'}\right|} 
\,\d^3 r' \, \d^3 r 
\nonumber\\ &&
+
 \frac{ Z^2 e^2}{2}\sum_{\stackrel{i,j=1}{i\ne
      j}}^N\frac{1}{\left| \mathbf{ R}_i-\mathbf{ R}_j \right|}
\end{eqnarray}
for a cluster of $N$ ions of valence $Z$ (for Na, $Z=1$),
valence-electron density $n$ and ionic coordinates $\{\mathbf{
R}\}$. The noninteracting kinetic energy $T_\mathrm{s}$ is calculated
from the Kohn-Sham orbitals $\{\varphi\}$ and $E_\mathrm{ xc}[n]$ denotes
the exchange and correlation energy for which we use the LDA
functional of Ref.\ \cite{pw}. Generalized gradient approximations,
e.g.\ \cite{pbe}, in general lead to a better description of
correlation effects in small systems. However, in the present case LDA
is not too bad an approximation due to the strong delocalization of
the valence electrons. $V_\mathrm{ ion}$ is the sum of pseudopotentials
$
  V_\mathrm{ ion}(\mathbf{ r; \{R\}}) =\sum_{i=1}^N v_\mathrm{
    ps}(\left|\mathbf{r - R}_i\right|).
$
We employ the smooth-core pseudopotential of Ref.\ \cite{locpp}. In
combination with LDA it provides accurate bond lengths, which are
important for optical absorption spectra and polarizabilities
\cite{locpp,therpol}. The ionic coordinates were obtained by
self-consistent minimization of the functional (\ref{clfunctional})
with respect to both $n$ and $\{\mathbf{R}\}$ and are given in Ref.\
\cite{locpp}. The Kohn-Sham equations are solved directly, i.e.,
without basis sets, on a real space grid. We have verified that the
coordinates obtained in the symmetry restricted optimizations of Ref.\
\cite{locpp} do not change noticeably if the optimization is done
fully three-dimensional. In order to be consistent, the ionic
configurations were reoptimized on the SIC level for the TDSIC
calculations, as discussed below.

Eq.\ (\ref{clfunctional}) is also the key ingredient for the quantum
fluid-dynamical LCA. A detailed discussion of the theory can be found
in Ref.\ \cite{lca}. Therefore, we here restrict ourselves to a brief
sketch. The essential idea of LCA is to describe excitations as
harmonic density oscillations. The oscillating density is obtained
from the scaling transformation
\begin{equation}
\label{scalen}
n({\mathbf r},\alpha(t))=e^{-\alpha(t) S_n} n({\mathbf r}),
\end{equation}
where $\alpha(t) \propto \cos \omega t$ and $S_n$ is the so called density
scaling operator 
\begin{equation}
\label{defsn}
S_n=\Big( \nabla \mathbf{u}({\mathbf r})\Big)+
\mathbf{u}({\mathbf r})\cdot\nabla,
\end{equation}
which contains -- hallmark of a fluid-dynamical description -- a
displacement field $\mathbf{u(r)}$.  A similar, consistent
transformation is also applied to the Kohn-Sham orbitals from which
the density is constructed. From a variational principle and Eq.\
(\ref{clfunctional}), a set of coupled, partial differential
eigenvalue equations for the Cartesian components of $\mathbf u$ is
derived. The eigenvalues are the excitation energies, and from the
solutions $\mathbf u_\nu$, absolute oscillator strengths and intrinsic
current densities
\begin{equation}
\label{defj}
  \mathbf{ j_\nu(r,} t)=\dot{\alpha}_\nu(t)\, \mathbf{ u_\nu( r}) \,
  n_\nu(\mathbf{r},\alpha_\nu(t))
\end{equation} 
associated with a particular (the $\nu$-th) excitation are derived.
Eq.\ (\ref{defj}) is the reason for the name ``local current
approximation''.  If a mode $\mathbf{ j_\nu}$ can be excited by the
dipole operator $D=-e z$ we call it a z-mode (or x,y, respectively).

It is important to note that the LCA is not a (semi) classical but a
quantum mechanical method in the sense that it is derived on the basis
of the quantum mechanical Kohn-Sham energy functional, which contains
information on the quantal single-particle states in the kinetic
energy. But the range of validity for the LCA is hard to assess
formally \cite{frompsitoe}. However, gathering experience on its
performance, as done for earlier versions which were truly
semiclassical methods \cite{brack89} or approaches using a well
guessed expansion basis of local operators \cite{lrpa}, will lead to a
better understanding. It is therefore one aim of the present work to
test the accuracy of the LCA by comparing it to the well established
TDLDA.

For TDLDA and TDSIC, the numerical solution of the Kohn-Sham equations
is done on a spatial grid with Fourier transformation for the
definition of the kinetic energy. Accelerated gradient iteration is
used for the static part. The dynamic propagation is done with the
time-splitting method.  For details of the technology see the review
\cite{ownrev}. The spectra are computed as described in
\cite{ownrev,YabBer,bigtdlda}. We start from the electronic and ionic
ground state configuration. An instantaneous boost of the whole
electron density initializes the dynamical evolution according to
time-dependent LDA or SIC. The emerging dipole momentum as function of
time is finally Fourier transformed into the frequency domain. This
delivers the spectral distribution of dipole strength. The initial boost is
kept small enough for the method to produce the spectra in the regime
of linear response.

\section{Results}
\label{results}

\begin{figure}
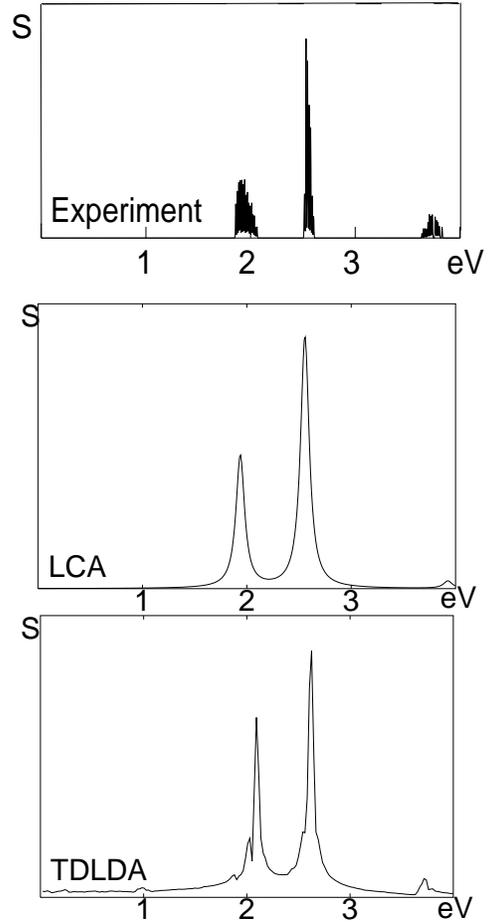

\begin{center}
\includegraphics[width=6.3cm]{na2nEXP.epsi}
\includegraphics[angle=270,width=6.cm]{na2nLCA.epsi}
\includegraphics[angle=270,width=6.cm]{na2nTDLDA.epsi}
\end{center}
\caption{Experimental \protect \cite{na2expdat}, LCA and TDLDA
photoabsorption spectrum S of $\mathrm{Na}_2$ in arbitrary units
against excitation energy in eV. The line broadening is chosen
phenomenologically to match the experiment.} 
\label{na2pabs}
\end{figure}

With the methods described in the previous section we first
investigated the sodium dimer. At first glance, it could be expected that
the two-electron system $\mathrm{Na}_2$ is not described accurately in
the quantum fluid-dynamical LCA. However, as seen from Fig.\
\ref{na2pabs}, TDLDA and LCA give similar results. Since the LCA
currents were calculated with the LDA functional, we first compare
them to TDLDA and discuss TDSIC results later. It is important to note
that our TDLDA and LCA calculations were performed on exactly the same
basis, i.e, using the same
\begin{table*}[hbt]
\caption{Dipole excitations up to 4 eV for $\mathrm{Na}_2$. Energies
EE in eV and oscillator strengths OS as percentages of the dipole
sumrule $m_1=e^2\hbar^2 N Z / (2m)$ for LCA (superscript $a$), TDLDA
(superscript b) and TDSIC (superscript c). Columns labeled ``mode''
indicate the direction of oscillation, see text for discussion. For
comparison, we also list TDLDA results (superscript d) and
experimental values (superscript e) from Ref.\ \cite{baerends}. ---
indicates that the corresponding mode is not found in LCA, - that the
strength in TDSIC was beyond numerical accuracy, no v. that no
corresponding value has been given in the literature.}
\label{tab1}
\begin{center}
\begin{tabular}{lllllllllll}
\hline\noalign{\smallskip}
EE$^a$  & OS$^a$ & Mode$^a$ &
EE$^b$  & OS$^b$ & Mode$^b$ &
EE$^c$  & OS$^c$ &
EE$^d$  & OS$^d$ &
EE$^e$             \\
\noalign{\smallskip}\hline\noalign{\smallskip}
 1.93 & 30.9 & z   & 2.09  & 31 & z   & 2.13 & 36
& 2.09 & 31.4 & 1.82 \\
 2.56 & 58.9 & x/y & 2.63  & 56 & x/y & 2.65 & 57
& 2.52 & 53.1 & 2.52 \\
 3.93 &  1.7 & z   & 3.67  & 1 & z   & 3.89  & -
& 3.28 & $<$1 & 3.64 \\
 ---  & ---  & --- & 3.72  & 3 & x/y & 3.95 & -
& no v.& no v.& no v.  \\ 
\noalign{\smallskip}\hline
\end{tabular}
\end{center}
\end{table*}
internuclear distance and pseudopotential. For a closer inspection we
give in Table \ref{tab1} the excitation energies and percentages on
the dipole sumrule for the excitations up to 4 eV that carry most of
the oscillator strength. Comparing columns 1--3 to columns 4--6
reveals some noteworthy differences between LCA and TDLDA. First, for
the lowest excitation, LCA gives an energy lower than TDLDA with a
difference larger than the numerical uncertainty. Since LCA rests on a
variational principle, the fact that it leads to a lower excitation
energy than TDLDA points at that it can be seen as an independent
method. In this context we also note that our TDLDA for the
z-excitation is consistent with the result in \cite{baerends}.
Second, whereas LCA is very accurate for the two low-lying, strong
transitions, it does not seem to perform as well for the higher lying
excitations. For technical reasons, the oscillator strength for the
weak transitions is hard to asses in our TDLDA and TDSIC and could
only be estimated in TDLDA. However, Fig.\ \ref{na2pabs} shows that in
comparison to TDLDA, the strength of the third peak is underestimated
in LCA, and the LCA eigenvalue is too high. Tab.\ \ref{tab1} shows why
the TDLDA spectrum in Fig.\ \ref{na2pabs} looks better. Whereas LCA
only leads to one z mode, TDLDA around 3.7 eV gives excitations in
both z and x/y direction. A ``double-peak'' structure has also been
found in other TDLDA calculations \cite{cheli}. Thus, we conclude that
LCA gives some of the strength carried by transitions at higher
energies, but it does not provide the same resolution as TDLDA. This
is understandable since TDLDA embraces the whole fragmentation into
the various one-particle-hole ($1ph$) states of the excitation
spectrum, while LCA is bound to a ``collective deformation path''.

For $\mathrm{Na}_2$, TDSIC leads to overall similar results as TDLDA,
therefore we do not show the spectrum in a separate plot. Besides the
fact that the TDSIC spectrum does not show the ``cut'' in the low
energy shoulder of the excitation at 2.09 eV, the main difference is
that TDSIC leads to slightly higher excitation energies than TDLDA
(see Tab. \ref{tab1}). The deviation is less than 0.05 eV for the
low-lying, strong transitions, but it is more than 0.2 eV for the
higher ones (placing them at a similar energy as LCA). The reason is
that the bonding distance is 0.1 a$_0$ smaller than in LDA
\cite{sic}. This slight compression leads to a small blue shift.  The
higher excitations are more sensitive because they are dominantly
$1ph$ transitions, and it is known that SIC has stronger effects on
the single-particle states. A recent, detailed discussion of how
several other approximations for $E_\mathrm{xc}$ influence the dimer
spectrum can be found in \cite{baerends}.

An intuitive understanding of the observed excitations can be obtained
by looking at ``snapshots of the density change''. In LCA, these are
easily accessible since the local currents, Eq.\ (\ref{defj}), obey the
continuity equation
\begin{equation}
\nabla \mathbf{j_\nu}+\frac{\d n_\nu(\mathbf{r},t)}{\d t}=0.
\end{equation} 
Thus, one only needs to numerically calculate and then plot the
divergence of $\mathbf{j_\nu}$ to obtain a visualization of the
density change associated with the $\nu$-th excitation at one
particular instant. This can be done separately for each LCA
eigenmode. As an analogon in TDLDA, we record the time evolution of the
density $n(\mathbf{r},t)$ and evaluate the Fourier components
\begin{equation}
\label{tdldatrans}
\tilde{n}(\mathbf{r},\omega_\nu)=
\int n(\mathbf{r},t) \exp (-i \omega_\nu t) \d t
\end{equation}
for the frequencies $\omega_\nu$ that are associated with particular
excitations. Since this procedure is numerically more demanding, we
have restricted the TDLDA analysis to a one dimensional section along
the axis of symmetry, integrating (\ref{tdldatrans}) over x and y
coordinates.

\begin{figure}[hbt]
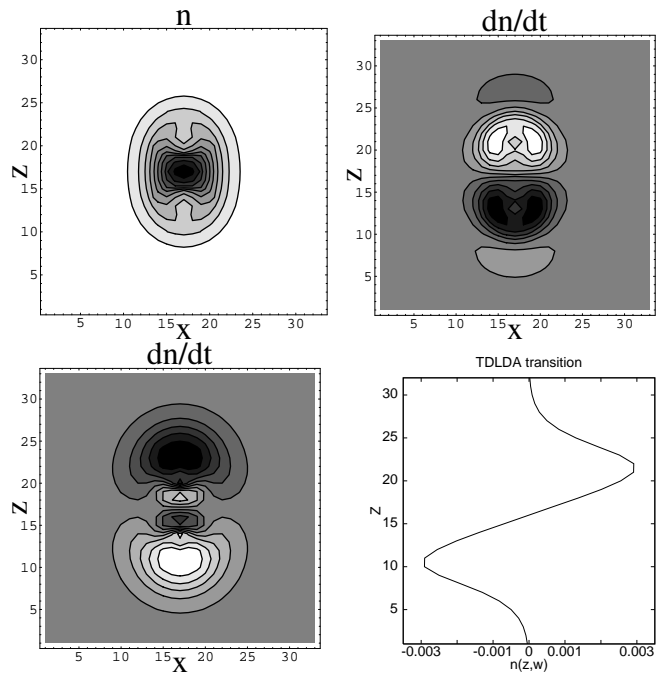

\includegraphics[height=4.4cm]{nGS.epsi}\hfill
\includegraphics[height=4.4cm]{dndtLCAmode10.epsi}\\
\includegraphics[height=4.4cm]{dndtLCAmode1.epsi}\hfill
\includegraphics[height=4.2cm]{na2TDLDAtrans.epsi}
\caption{``Snapshots'' of the changes in valence electron density
associated with particular excitations of $\mathrm{Na}_2$. Contour 
plots show a plane containing the axis of symmetry of the
molecule. Unit for the axes is the numerical grid spacing, 0.8
$a_0$. Top left: ground-state valence electron density $n$. Bottom
left: $\d n /\d t$ associated with the first excitation, i.e. first 
z-mode, in LCA. Top right: $\d n /\d t$ associated
with the third excitation, i.e. second z-mode, in
LCA. Shadings lighter than the background gray indicate a density
decrease, darker shadings an increase. Bottom right:
$\tilde{n}(\mathbf{r},\omega_\nu)$ integrated over x and y as a
function of z for lowest (i.e., $\nu=1$) TDLDA excitation. The
pictures are in accordance with understanding the excitations as
density oscillations (see text).} 
\label{dndt}
\end{figure}
The top left picture in Fig.\ \ref{dndt} shows a contour plot of the
ground-state valence electron density of $\mathrm{Na}_2$ in a plane
containing the axis of symmetry. (The grid for the calculation was
larger than the shown part.) The ionic cores are clearly visible since
they repel the valence electrons, leading to ``holes'' in the
density. The bottom left picture visualizes how this valence electron
density changes in time at the first excitation. Dark colors indicate
a density increase, light colors a decrease. Obviously, the electron
density increases at one end of the molecule and decreases at the
other end. Thus, the valence electrons are shifted predominantly along
the axis of symmetry. But the shift is not a uniform, simple
translation of the density along the dipole field (as it would be
obtained from the sumrule estimate \cite{revmod}), but the intrinsic
structure of the cluster is impressed on and reflected in the
currents, leading to a shift ``around'' the ionic cores. The second
excitation in LCA is a (twofold degenerate) x/y mode. Its density
change (not shown in Fig.\ \ref{dndt}) is predominantly perpendicular
to the axis of symmetry, as one also naively would expect. The third
LCA excitation, shown in the top right picture, is again a z mode. The
regions of strongest density variation are shifted compared to the
first z mode, and the oscillation pattern shows a node at greater
separation from the ionic cores. This reflects the mathematical
requirement of orthogonality for the different modes
\cite{brack89,lca,diss},
\begin{equation}
\label{orthogonality}
\int \mathbf{u}_\nu(\mathbf{r}) \mathbf{u}_\mu(\mathbf{r}) 
\, n(\mathbf{r}) \, \d^3r \propto \delta_{\mu\nu}.
\end{equation}
Physically, the plots show that the observed electronic transitions
can be interpreted as different eigenmodes, i.e., intrinsic
oscillation patterns of the valence electron distribution. The TDLDA
transition densities confirm this picture. The bottom right part of
Fig.\ \ref{dndt} shows Eq.\ (\ref{tdldatrans}) evaluated for the
frequency of the lowest mode. Since x and y coordinate have been
integrated over, some of the finer structure might have been smoothed
out. But it is clearly visible that also in TDLDA, the lowest
excitation is associated with a density increase at one end of the
molecule and a decrease at the other. The regions of maximum density
change are found very similar in LCA and TDLDA. We have verified this
also for the second excitation that is not shown in Fig.\ \ref{dndt}.
Thus, the simple picture of excitations as density oscillations seems
to be remarkably close to the truth, even for the small
$\mathrm{Na}_2$.

\begin{figure}
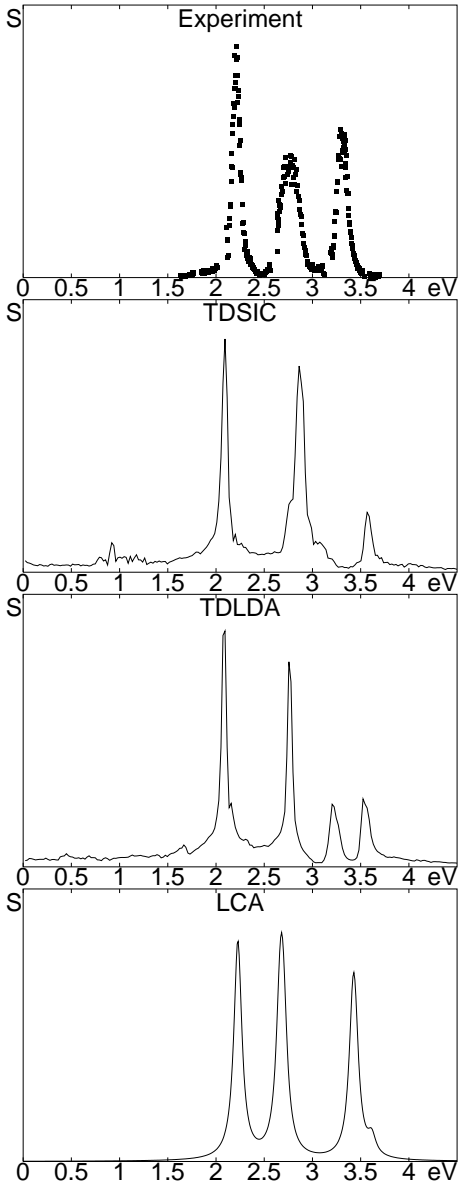

\begin{center}
\includegraphics[angle=270,width=6.cm]{na5pEXP.epsi}
\includegraphics[angle=270,width=6.cm]{na5pTDSIC.epsi}
\includegraphics[angle=270,width=6.cm]{na5pTDLDA.epsi}
\includegraphics[angle=270,width=6.cm]{na5pLCA.epsi}
\end{center}
\caption{From top to bottom: Experimental \protect \cite{charpabs},
TDSIC, TDLDA and LCA photoabsorption spectrum S of $\mathrm{Na}_5^+$ in
arbitrary units against excitation energy in eV.} 
\label{na5pabs}
\end{figure}

Fig.\ \ref{na5pabs} shows the experimental low-temperature
photoabsorption spectrum \cite{charpabs} of $\mathrm{Na}_5^+$ and
below the spectra obtained in TDSIC, TDLDA, and LCA.
Again, the LCA results (bottom) are shown with a phenomenological line
broadening to make comparison with TDLDA easier. Overall, the spectrum
obtained in LCA is rather close to the experiment, which again is
remarkable in view of the small cluster size. However, our main focus
here is on the comparison between TDLDA and TDSIC. TDLDA gives the
energies and relative peak heights for the two lower transitions close
to the experimental ones. But instead of one peak that is seen
experimentally at about 3.3 eV, TDLDA leads to two peaks at 3.20 eV
and 3.53 eV. A similar pattern was also found but not explained in
recent TDLDA calculations \cite{moseler} that focused on the
explanation of the observed linewidths. By going over to TDSIC, we can
further investigate the nature of the double peak. The TDSIC spectrum
in Fig.\ \ref{na5pabs} shows noticeable differences to TDLDA. First, a
small subpeak is found at 0.92 eV. It is a relic of a $1ph$ excitation
which originally was close to 1 eV and which has given most of its
strength to the dominant peak. The comparable state in TDLDA is found
at about 1.5 eV, i.e., so close to the main peak that its strength is
hardly recognizable. Second, whereas the peak at 2.09 eV is hardly
changed by the averaged SIC, the peak which in TDLDA was at 2.76 eV
shifts to 2.86 eV in TDSIC and appears broader since there is another
transition close by at 3.03 eV. This might contribute to explaining
why also in the experimental low temperature data, the middle peak
appears to be somewhat broader. Finally, the last excitation again
stays nearly unchanged at 3.57 eV. Thus, by shifting the peak which in
our TDLDA is found at 3.20 eV to lower energies, TDSIC leads to a
spectrum that is close to the experimental one. We also tested whether
this is only an indirect effect, due to slight rearrangements when the
ionic geometry is reoptimized on SIC level. However, this is not the
case: even when compared for exactly the same ionic structure, TDLDA
and TDSIC spectra show noticeable differences. We find, in accordance
with earlier investigations \cite{sic}, that TDSIC leaves the main
resonance peaks basically unchanged, but it strongly modifies the
single-particle energies, and thus the underlying $1ph$ excitation
spectrum.  Our comparison with experimental data shows that while
TDLDA gives reasonable results for the gross features of a spectrum,
it can be inaccurate for details. In the energy range and for the
clusters studied here, the TDSIC description improves on TDLDA
deficiencies in details of the coupling to $1ph$ structures.

\section{Conclusions}
\label{conclusions}

Our investigation of the photoabsorption spectra of two small sodium
clusters with three different methods shed new light on the
theoretical methods as well as on the understanding of the
experiments. The local current approximation is based on a
``collective'' picture of excitations. It exploits information that is
contained in the curvature of the ground-state energy functional, and
its success in the cases studied here demonstrates that the
functional, indeed, contains relevant information on the excited
states. Furthermore, LCA's quantitatively accurate description of the
strong excitations and partial success in describing higher lying ones
shows that the concept of ``collectivity'' and the detailed view in
terms of particle-hole excitations have more in common than expected,
even for these small clusters. TDLDA has the advantage of being robust
throughout a wide range of energies. It leads to a reliable
description of overall features of photoabsorption spectra. The
comparison with TDSIC showed, however, that details of the excitation
spectrum can be rather sensitive to self-interaction effects, even for
a simple metal like sodium. Thus, treating exchange and correlation on
a level beyond LDA is very desirable.

The success of density-functional methods to accurately describe the
excitation spectra of small clusters in general, and the LCA in
particular, leads to an alternative interpretation of the
photoabsorption data. Our results show that the traditional way of
thinking of excitations in small metal clusters as transitions between
distinct molecular states coincides nicely with the more intuitive way
of understanding them as collective electronic eigenmodes, i.e.,
oscillations of the valence electron density. Of course, these
oscillations are neither exactly like the Mie plasmon in a classical
metal sphere nor like the compressional bulk plasmon. Their
frequencies and oscillation patterns are determined by the clusters'
intrinsic structure, which for small systems like the ones studied
here must of course be described quantum mechanically. But if this is
taken into account, the picture of density oscillations is well
compatible with the ``molecular states'' point of view, and with
experimental data.

\begin{acknowledgement}S.K.\ acknowledges discussions with M.\ Brack
and financial support from the Deutsche
Forschungsgemeinschaft under an Emmy-Noether grant. 
\end{acknowledgement}


\begin{thebibliography}{}

\bibitem{neupabs}C.\ R.\ C.\ Wang, S.\ Pollack, D.\ Cameron, and M.\
  M.\ Kappes, Chem.\ Phys.\ Lett.\ \textbf{ 166}, 26 (1990); J.\
 Chem.\ Phys.\ \textbf{ 93}, 3787 (1990). 

\bibitem{selby}K.\ Selby, V.\ Kresin, J.\ Masui, M.\ Vollmer, W.\ A.\ 
  de Heer, A.\ Scheidemann, and W.\ D.\ Knight, Phys.\ Rev.\ B \textbf{
    43}, 4565 (1991).

\bibitem{meiwesbroer}For a recent example, see e.g. T.\ Doppner, S.\
Teuber, M.\ Schumacher, J.\ Tiggesbaumker, K.H.\ Meiwes-Broer, Appl.\
Phys.\ B \textbf{71}, 357 (2000). 

\bibitem{charpabs}C.\ Ellert, M.\ Schmidt, C.\ Schmitt, T.\ 
  Reiners, and H.\ Haberland, Phys.\ Rev.\ Lett.\ \textbf{ 75}, 1731
(1995); M.\ Schmidt and H.\ Haberland, Eur.\
  Phys.\ J.\ D \textbf{ 6}, 109 (1999). 

\bibitem{ep}W.\ Ekardt, Phys.\ Rev.\ B \textbf{31}, 6360 (1985); W.\ Ekardt
and Z.\ Penzar, Phys.\ Rev.\ B \textbf{ 43}, 1322 (1991).

\bibitem{brack89}M.\ Brack, Phys.\ Rev.\ B \textbf{ 39}, 3533 (1989).

\bibitem{lrpa}P.-G.\ Reinhard, M.\ Brack and O.\ Genzken, Phys.\ 
  Rev.\ A \textbf{41}, 5568 (1990).

\bibitem{guet}M.\ Madjet, C.\ Guet, and W.\ R.\ Johnson, Phys.\ Rev.\
A \textbf{ 51}, 1327 (1995).

\bibitem{schmidt}U.\ Saalmann and R.\ Schmidt, Z.\ Phys.\ D \textbf{
38}, 153 (1996).

\bibitem{rubio}A.\ Rubio,
  J.\ A.\ Alonso, X.\ Blase, L.\ C.\ Balb\'{a}s, and S.\ G.\ Louie,
  Phys.\ Rev.\ Lett.\ \textbf{ 77}, 247 (1996); M.A.L.\ Marques, A.\
  Castro, and A.\ Rubio, J.\ Chem.\ Phys.\ \textbf{115}, 3006 (2001). 

\bibitem{ci}V.\ Bona\v{c}ic-Kouteck\'{y}, J.\ Pittner, C.\ 
  Fuchs, P.\ Fantucci, M.\ F.\ Guest, and J.\ Kouteck\'{y}, J.\ Chem.\ 
  Phys.\ \textbf{ 104}, 1427 (1996).

\bibitem{sic}
C.\ A.\ Ullrich, P.-G.\ Reinhard, and E.\ Suraud, Phys.\ Rev.\ A
\textbf{62}, 053202 (2000).

\bibitem{cheli}I. Vasiliev,
S. \"O\u{g}\"ut, and J. R. Chelikowsky, Phys. Rev. Lett. \textbf{ 82},
1919 (1999).

\bibitem{locpp}S.\ K\"ummel, M.\ Brack, and P.-G.\ Reinhard,
  Phys.\ Rev.\ B \textbf{62}, 7602 (2000).

\bibitem{baerends}S.\ J.\ A.\ Gisbergen, J.\ M.\ Pacheco, and E.\ J.\
Baerends, Phys.\ Rev.\ A \textbf{63} 062301 (2001).

\bibitem{tddft}For overviews see, e.g., E.\ K.\ U.\ Gross, C.\ A.\ Ullrich,
  and U.\ J.\ Gossmann, in \textit{Density Functional Theory}, edited by E.\
  K.\ U.\ Gross and R.\ M.\ Dreizler (NATO ASI series, Plenum, New
  York 1994); E.\ K.\ U.\ Gross, J.\ F.\ Dobson, and M.\ Petersilka,
  in {\it Density Functional Theory}, edited by R.\ F.\ Nalewajski
  (Topics in Current Chemistry, Vol.\ 181, Springer, Berlin, 1996);
  for interesting recent caveats see N.T.\ Maitra and K.\
  Burke, Phys.\ Rev.\ A \textbf{63}, 042501 (2001).

\bibitem{lca}S.\ K\"ummel and M.\ Brack, Phys.\ Rev.\ A \textbf{64},
  022506 (2001).

\bibitem{pw}J.\ P.\ Perdew and Y.\ Wang, Phys.\ Rev.\ B \textbf{45},
  13244 (1992).

\bibitem{pbe}J.\ P.\ Perdew, K.\ Burke, and M.\ Ernzerhof, Phys.\
  Rev.\ Lett.\ \textbf{77}, 3865 (1996).

\bibitem{therpol}S.\ K\"ummel, J.\ Akola, and M.\ Manninen,
  Phys.\ Rev.\ Lett.\  \textbf{84}, 3827 (2000).

\bibitem{frompsitoe}At the heart of this matter is Eq.\ (27) of Ref.\
\cite{lca}. 

\bibitem{ownrev}
F.\ Calvayrac, P.-G.\ Reinhard, E.\ Suraud, C.\ Ullrich,
Phys.\ Rep.\ \textbf{ 337}, 493 (2000).

\bibitem{YabBer}
K.\ Yabana, G.F.\ Bertsch, Z.\ Phys.\ D \textbf{ 42}, 219 (1997).

\bibitem{bigtdlda}
F.\ Calvayrac, E.\ Suraud, P.-G.\ Reinhard,
Ann.\ Phys.\ \textbf{ 254} (N.Y.), 125 (1997).

\bibitem{na2expdat}W.\ R.\ Fredrickson and W.\ W.\ Watson, Phys.\ Rev.\
  \textbf{30}, 429 (1927); presentation of experimental data adapted from
  Ref.\ \cite{cheli}. 

\bibitem{revmod}An extensive discussion of the sumrule approach can be
found, e.g., in M.\ Brack, Rev.\ Mod.\ Phys.\ \textbf{65}, 667 (1993). 

\bibitem{diss}S.\ K\"ummel, \textit{Structural and  Optical Properties
of Sodium Clusters studied in Density Functional Theory} (Logos
Verlag, Berlin, 2000) 37. 

\bibitem{moseler}M.\ Moseler, H.\ H\"akkinen, U.\ Landman, Phys.\
Rev.\ Lett.\  \textbf{87}, 053401 (2001).


\end{thebibliography}
\end{document}